# HANDLING MOBILITY IN DENSE NETWORKS

Ngọc-Dũng Đào, Hang Zhang, Hamid Farmanbar, and Xu Li
Huawei Canada Research Centre,
Unit 400, 303 Terry Fox Drive, Ottawa, Canada, K2K 3J1
Email: {Ngoc.Dao, Hang.Zhang, Hamid.Farmanbar, Xu.LiCA}@huawei.com

**Abstract**: Network densification is one of key technologies in future networks to significantly increase network capacity. The gain obtained by network densification for fixed terminals have been studied and proved. However for mobility users, there are a number of issues, such as more frequent handover, packet loss due to high mobility, interference management and so on. The conventional solutions are to handover high speed mobiles to macro base stations or multicast traffic to multiple base stations. These solutions fail to exploit the capacity of dense networks and overuse the backhaul capacity. In this paper we propose a set of solutions to systematically solve the technical challenges of mobile dense networks. We introduce network architecture together with data transmission protocols to support mobile users. A software-defined protocol (SDP) concept is presented so that combinations of transport protocols and physical layer functions can be optimized and triggered on demand. Our solutions can significantly boost performance of dense networks and simplify the packet handling process. Importantly, the gain brought by network densification to fixed users can also be achieved for mobile users.

## 1 Introduction

The network densification is among technology candidates for the fifth generation (5G) radio access networks (RAN) [1]. Densely and ultra-densely deployed networks could become main deployment scenarios in future. In dense networks (DN), the distance between radio nodes could be tens of meters, which is much smaller than that found in the today macro networks. While the DN can provide very large throughput, it also poses significant technical challenges [1][15]. First, because of such small inter-cell distance, the number of handover events could happen very often for mobility users. Second, the co-channel interference may become excessive because of large number of cells, especially when they are deployed unplanned.

The two problems in DN have profound impacts to the performance of transport layer protocol, such as popular TCP [1] and UDP [3] protocols. For example, the TCP source relies on packet round-trip-time (RTT) to adjust the sending rate. When the wireless channel capacity varies quickly as in DN mobility scenario, the RTT estimation is no longer reliable and the end-to-end network performance could dramatically drop. The great potential capacity of DN could be hardly harvested with the today transport protocols and conventional macro-cell network management.

Conventional solution to avoid problems in the mobile DN is to associate high mobility users to macro cells. Because of the large cell sizes, the number of handover will be reduced. Another solution is multicast, where the same data is sent to multiple radio nodes. The schedulers are tightly coordinated to transmit data to users. However, the first solution will not take advantage of large capacity in DN. The second solution uses more bandwidth in the backhaul, and requires highly complex transmission coordination.

The software-defined network (SDN) concept is a strong candidate to provide global network management and operation solutions [13][14]. It promises an end-to-end solution for traffic management, taking into account dynamic traffic demands, quality of experience (QoE) requirement, and network capacity. In the SDN framework, network control functions are separated from data forwarding functions. For example, traffic engineering (TE), which optimizes the routing and resource allocation, can be taken out of routers; routers can simply perform packet forwarding based on routing decision made by the TE. In case of DN, the SDN framework can be applied to balance the traffic load among nodes and configure the resources such as frequency and transmit power. Hence the interference can be regulated [11][12].

Anyway, introduction of SDN alone may be still not sufficient to handle variety of traffics with distinguished QoE requirements. Additional fundamental network management tools need to be considered. In current networks, different network layers provide their own measures for reliable packet delivery. TCP protocol has automatic retransmission; physical layer provides hybrid automatic retransmission; handover procedure may forward packets to new serving cells. These three mechanisms, if not properly configured, actually cause packet duplication and deteriorate the capacity of DN, while adding unnecessary complexity.

In this article, we present solutions for some of most challenging problems in the DN, as well as solutions for mobility handling in this network. In particular, we propose the following solutions.

- Network architecture to support mobility and interference management in DN networks.
- Software-defined protocol (SDP) concept that optimizes packet handling procedures across network layers, including transport layer protocols, packet handover protocol, and physical layer functions for individual traffic types in mobility.

Our results show that the application of SDP can efficiently utilize the capability of SDN framework, significantly improve the performance of DN, and optimize packet forwarding procedures. Simulation results prove that the gain brought by network densification to fixed users can also be achieved for mobile users.

## 2  Network Architecture

A simplified network model is illustrated in Figure 1. A mobile user requests data files from a media server while travelling across the network. The user equipment (UE) has a logical connection to radio access network (RAN) through one or more physical radio nodes, e.g. WiFi access point and cellular radio nodes, at the same time.

When users are moving, the capacity of wireless channels can significantly vary. To catch up with the channel variation, the TE Optimizer may need to run more frequently. However, frequent runs of TE may require large feedback information from wireless users and flow update commands to network routers. Consequently, communication overhead and optimization complexity would place a certain limit on how often a global TE optimization may be performed.

One solution to reduce the number of TE optimization runs is network segmentation. The whole network can be divided into two segments, namely, fixed and user-specific network segments. A virtual user-specific serving gateway (v-u-SGW) is dynamically placed in the border of two network segments. The wireless links belong to the user-based mobile segment. The user-specific mobile segment may also include a number of wireline hops from radio nodes to the v-u-SGW. The fixed network segment consists of only stable wired links. There will be two separate TE runs for each network segment. A longer timescale TE solution can be applied to the fixed network segment; this TE is able to handle very large number of users, which represent by v-u-SGWs. The second TE handles traffic from the v-u-SGW to the user devices. Since the number of users managed by a virtual v-u-SGW is small, the TE may be run more frequently.

The v-u-SGW can perform additional control functions, such as authentication, as well as data forwarding functions. In particular, the v-u-SGW can be equipped with caching capability so that it can resend missing packets. The v-u-SGW can also generate repair packets, for example fountain coded packets, for faster error correction. The v-u-SGW thus can significantly reduce the end-to-end file transmission delay and network management complexity.

Next we discuss a TE solution and v-u-SGW placement optimization.

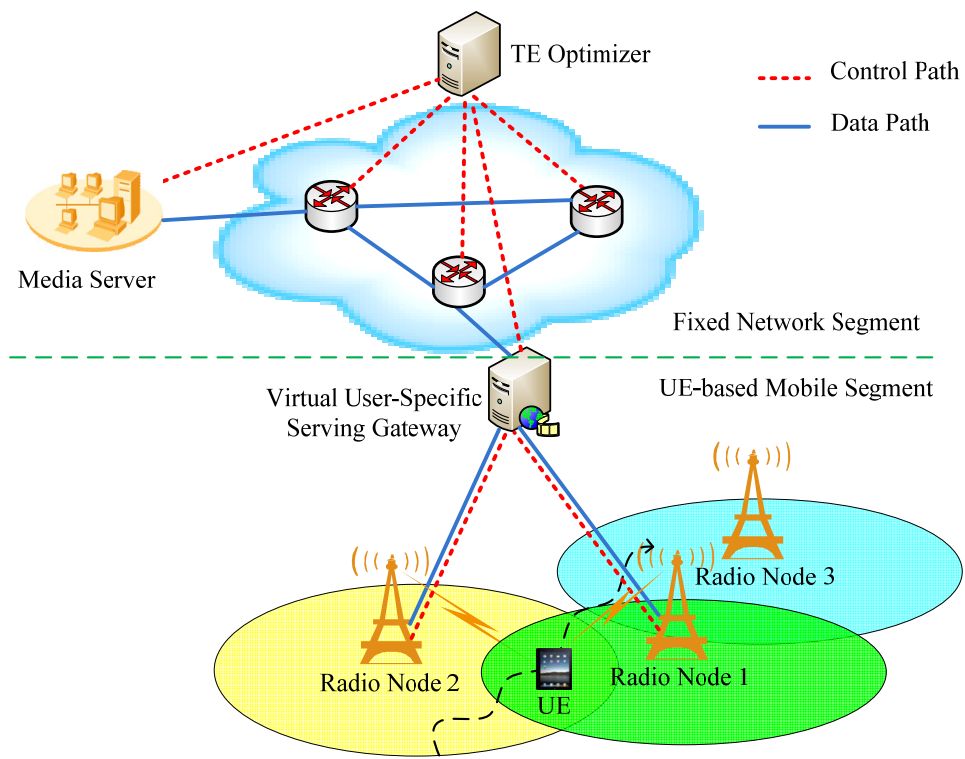

**Figure 1: Simplified network model to support mobile users.**

## 2.1 Traffic Engineering

The TE function deals with routing flows from their sources to their corresponding destinations. A flow is identified by a source-destination pair. A flow often has a demand, e.g., rate requirement. The problem of routing flows in a network given flow demand requirements and network link capacity constraints is called multi-commodity flow problem, where a commodity refers to a flow along with its corresponding demand.

In the multi-path TE, several paths are used for each flow to satisfy flow demand. Compared to the single-path TE, the multi-path TE provides more flexibility for flow routing as there are multiple paths available for each flow. In the context of software-defined networks (SDN), the TE function is implemented in a logically *central controller* which determines a set of paths as well as rate allocation on each path for each flow based on the knowledge of network topology, link capacities, and flow demands.

What makes TE different in a radio access network (RAN) compared to wired network TE is that the capacity of a wireless link in a radio access network is not fixed and depends on the amount of resource (e.g., frequency, time, etc.) assigned to it by the access link scheduler. This can be dealt with by *abstraction* of radio access link scheduling. Radio access link scheduling abstraction describes the relationship between the supported rates from access points to UEs and the network radio resources. The outcome of abstraction is a rate region that can be supported by the radio access links. Such wireless rate region defines the trade-offs among supported data rates on wireless links connecting the radio access nodes to their users. When such an abstraction is available, the SD-RAN TE problem becomes similar to wired network TE problem with additional wireless rate region constraints [12].

In the path-based TE approach, a number of candidate paths for all source-destination pairs are selected and traffic allocation for all commodities is performed over those paths. A special case is the single-path scenario where there is only one candidate path per commodity. In the path-based approach, the variables of multi-commodity flow problem are flow allocation on paths. The constraints of the multi-commodity flow problem include *flow demand constraints*, *backhaul link constraints*, and w*ireless access node constraints*.

The traffic splitting decisions for flows are obtained by solving an optimization problem subject to the above constraints. The objective function of the optimization problem is selected based on network operator's requirements in terms of load balancing, over-provisioning, and resource usage control in the network.

## 2.2 Virtual User-Specific Serving Gateway Placement

As elaborated earlier, v-u-SGW is a solution to reducing the impact of UE mobility to the complexity of network management and resource allocation. Assume that a set of network nodes is pre-configured to be v-u-SGW host. Compared to regular nodes, they have large bandwidth, high data processing power, and large storage space to accommodate massive traffic and engage mobility management. Given the host set, it is a challenge to select a subset of them and place v-u-SGWs on the selected subset in such a way the v-u-SGWs effectively mask UE mobility and best leverage network performance, TE efficiency and stability at minimal control overhead.

It may be assumed that hop count is used or considered as part of the routing criteria. The routing path length (hop count) between a source and a destination is in general proportional to the physical distance between them. Consider a routing path from a source to a destination UE. When the UE moves, the path may change, and the change occurs in a segment of the path adjacent to the destination. The changing segment will occupy a large portion of the path if the destination mobility is high and a small portion otherwise. This is the so-called distance effect. The v-u-SGW is insensitive to the UE's motion, that is to say, masking it from the source, if it is outside the changing segment. If it is placed close to the source, the chance is high and low otherwise.

If a v-u-SGW is close to the source and far from the UE, network resources may not be best utilized (for example, when traffic inflation occurs between the v-u-SGW and respective UE), and TE complexity is not well controlled. If, on the other hand, it is near the UE and far from the source, it will not mask UE mobility well and will cause frequent v-u-SGW migration (i.e. replacement), unstable TE decision, and therefore increased control overhead. A network node may host multiple v-u-SGWs. The number of selected v-u-SGW hosts should be kept as small as possible for controlling operating cost. A good trade-off needs to be found during v-u-SGW placement.

### 2.3  Interference Management

Low power nodes in DN have small coverage range and thus may serve only a few users. The user and traffic distributions in small cells could be highly heterogeneous. Additionally, being deployed closely, small cells may cause excessive interference to neighbor cells. Therefore, interference management is crucial to provide high data rates, especially for mobile users when they cross cell borders.

We have proposed an on-demand radio coordination framework for DN in [11]. In this framework, the interaction of TE Optimizer and a RAN Optimizer is described. The TE Optimizer provides traffic load information, and the RAN Optimizer dynamically configured the radio nodes parameters to meet the traffic demands. The RAN Optimizer can create clusters of radio nodes surrounding congested nodes to reduce the optimization complexity as well as efficient load balancing. The proposed on-demand RAN coordination can effectively increase the capacity of DN [11].

## 3  Software-Defined Protocol

Supporting mobility in DN requires fundamental changes to the way data packets are handled in transport layer as well as in physical layer. In this section, we will discuss technical issues of current transport layer protocols and physical layer, together with a potential solution, namely software-defined protocol (SDP). The SDP provides network-wide coordination to optimize packet handling protocols at different layer for different traffic types.

The most popular transport layer protocol in the Internet is TCP [2]. TCP provides functions for reliable delivery, rate control and congestion avoidance. In SDN networks, the routing and rate allocation are performed by TE. Therefore, these two functions of TCP can be eliminated; only the automatic packet retransmission function could be kept. We call this protocol as TCP-Decomposition (TCP-D) in this study.

In DN, frequent handover is unavoidable as users may spend a few seconds in small cells. The handover process involves control plane and data plane procedures. In this paper, we concentrate on the data plane packet handling functions only. There are two packet processing modes in cellular networks at handover, packet dropping and packet forwarding [8]. In packet dropping mode, packets are simply dropped in the current serving cells at handover. On the other hand, in packet forwarding handover procedure, the current serving cell forwards unsent packets to new and other serving cells. This procedure requires packet forwarding from the current serving radio node to the v-u-SGW and then v-u-SGW will forward packets to the other serving cells.

The packet forwarding handover seems to provide better packet delivery at the first glance. However, it may cause packet duplication problem in TCP sessions. Since the capacity of wireless channel in DN can quickly vary because of mobility, the packet delay is also varying. Hence the TCP sender may be unable to reliably estimate the packet delay for automatic packet retransmission based on packet round-trip-time (RTT). The TCP source may resend the same packets to the serving cells and cause packet duplication. This issue is more severe when UEs are in cell edges, where handover occurs, and the lower channel capacity causes longer packet delay.

In cellular systems, physical layer also provides packet retransmission, for example hybrid automatic request retransmission (HARQ) [8]. In radio nodes, the packet scheduler selects data rate based on measurement feedback from user devices. If the measurement is imprecise due to channel changing and interference variation as in mobility scenario, the packet transmission is subject to errors. In this case, radio nodes may retransmit the packets using HARQ. However, the HARQ retransmissions may cause certain additional delay. In long-term evolution (LTE) systems, the minimum delay is 8 ms [8]. The TCP automatic retransmission may be triggered and thus cause packet duplication. Also handling packets in HARQ introduces non-negligible complexity to the physical layer.

So, different layers in the network have their own packet protection mechanisms. These mechanisms can be conflicting and bring unnecessary complexity to the system operation.

## Fountain Coding

One promising solution to avoid packet duplication is to employ fountain codes (FC) (also known as rateless coding) [4][5][6] in combination with multipath TE. This protocol is called fountain coded multipath transmission (FC-MP) protocol. The most outstanding feature of FC is that the FC encoder can create practically infinite number of independently encoded packets from a finite number of data packets with very low encoding and decoding complexity. Receivers just need to collect a certain number of arbitrarily coded packets in order to decode the whole message. Latest RaptorQ codes [6] require very small coding overhead. Fountain codes have been standardized in many Internet file transfer protocols, for example in multimedia broadcast multicast system (MBMS) of 3GPP cellular networks [7].

In our solution, the v-u-SGW collects data packets of a flow and use systematic FC to create coded packets. The coded packets are sent over multiple paths to multiple serving radio nodes. Radio nodes perform independent packet scheduling and transmit FC packets to the receiver until the receiver can decode the original data file. The data rates on multiple paths of multiple flows are jointly optimized by the TE Optimizer, taking into account the capacity of wired and wireless links.

## Fountain Coding for Real-Time Traffic

For real-time services, the end-to-end packet delay is very stringent, 100 ms for real-time video conferencing for example [8]. The transmission protocol for real-time video should be very different from that of BE traffics. Real-time video systems often employ UDP protocol [3] in transport layer since video decoder can tolerate a certain packet loss rate. The missing video packets could be fixed by error concealment techniques in video decoders. However, the video viewers may still recognize erroneous blocks of pixels where original bits are replaced by error concealed bits.

In most video distribution systems, video frames of group of pictures are jointly encoded [10]. This causes encoding and decoding delays of one or few seconds, which is unacceptable for interactive video services. To avoid the encoding/decoding delays, we propose to protect single video frames by fountain coding as illustrated in Figure 2. The v-u-SGW generates fixed-rate FC for I-frames and distributes coded packets to multiple radio nodes. The fixed rate FC means that the ratio of the original packets and coded packets is fixed. There is no packet scheduling coordination among radio nodes. At the receiver, once a fountain coded I-frame is successfully decoded, the receiver will inform radio nodes to remove remaining packets of this I-frame to save air interface resources.

The FC-MP protocol can avoid packet duplication of the TCP protocol. We will further investigate whether packet forwarding handover and HARQ are still needed for popular BE and delay-sensitive video traffics by simulations in the next section. Simulation results in a complex network environment will help to understand when and where network functions need to be triggered within SDP concept.

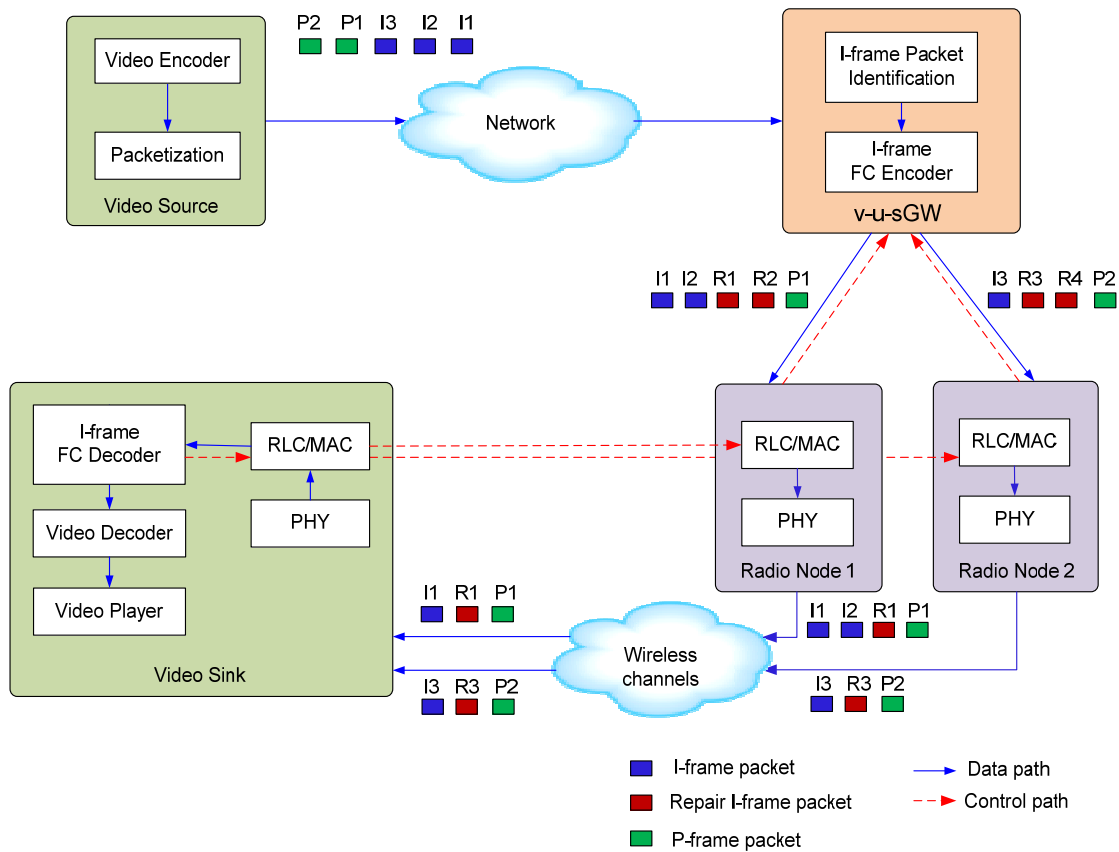

Figure 2: System model to apply fountain coding for real-time video.

# 4 Network Performance Evaluation

## 4.1 Simulation Settings

**Network Model**

The simulated network topology serving downlink traffics is plotted in Figure 3. It is a dense network of 57 radio nodes, connected to 11 routers in the core network. The radio nodes are randomly deployed in an area of 0.04 km$^2$. Three out of eleven core routers are connected to a gateway. For this small-sized network, the v-u-SGW is located in the network gateway. The traffic flows are initiated from the gateway and routed to mobile users. Operation of radio nodes follows LTE specifications with 10 MHz bandwidth and 2-by-2 MIMO closed-loop modes. The transmitter and receiver antennas are omni-directional.

## Mobility Model

A highway mobility scenario is considered, where users travel in two parallel straight lines through the network. Users monitor the downlink path losses and report the path losses of the best N serving cells.

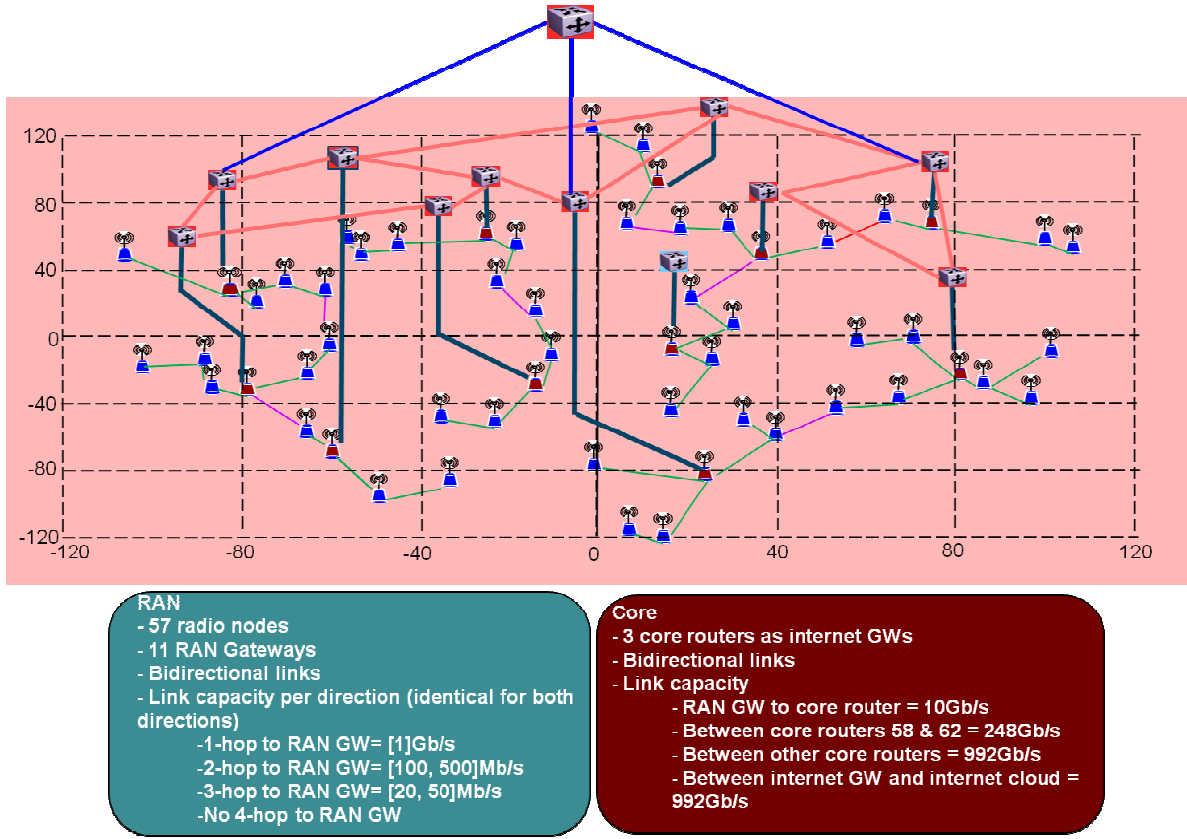

**Figure 3: Simulated network topology.**

## Traffic Models

There are two traffic models, best-effort (BE) and real-time video. In BE sessions, users download files of the same size 20 Mbit. The off-time between two sessions follows an exponential distribution with a given mean off-time, which is set to 10 seconds (low intensity) and 1 second (high intensity). The number of completed sessions is used as the performance metric.

For real-time video player, like video conferencing, packet delay plays an important role in the video decoding quality. At the viewing time, if some or all packets of a video packet are missing, the viewing quality for this video frame is degraded. The total time the video quality degraded (outage time) is one of key metrics indicating the network performance. In our evaluation, the video rate at which 99% of video sessions having less than 5% outage time is used for comparing performance of transport protocols.

In terms of TE, the difference between BE and real-time video is that the network needs to provide a stable data rate for real-time video, while the assigned rate for BE traffic can vary depend on users' channel capacity. The TE Optimizer will assign the data rate to BE flows based on max-sum throughput algorithm, or max-min throughput for video flows. When users move to the cell borders, the data rate could be significantly reduced. Thus for video flows, a multi-cell resource coordination, such as power control, is required to improve the cell-edge performance [11].

## 4.2 Performance of Best-Effort Traffic

To evaluate performance of FC protocol, which heavily depends on the optimality of TE, we assume that the backhaul capacity is significantly larger than that of wireless channels. In this case, the v-u-SGW could send coded packets at the same high rate (20 Mbps) to multiple serving radio nodes. Radio nodes perform independent packet scheduling to transmit coded packets the receiver. However, due to different wireless channel capacity, radio nodes may send coded packets at different rates to a single receiver. We call this protocol is fountain coded multicast (FC-MC) protocol. The performance of FC-MP protocol will be compared to the FC-MC protocol. The high rate of FC-MC flows will quickly fill up the transmit buffer of radio nodes to so that radio nodes always have data to transmit to users whenever users are scheduled. The FC-MC protocol would be the benchmark for FC protocol if backhaul capacity is practically large.

Simulation results for BE traffic are presented in Figure 4, where performance of TCP-D, FC-MP, and FC-MC is compared to that of single-path TCP-D protocol. There are 30 users, moving at 30 km/h. Packets are dropped when users move to new serving cells. We will verify the impacts of packet forwarding handover later. The major observations are as follows.

- The FC-MP protocol provides considerably higher gains compared to TCP-D protocol. There is a certain gap to the upper bound performance of FC-MC protocol due to inaccurate channel information. We will later discuss a technique to improve TE performance in mobility scenario.

- The role of multipath TE is more dominant if traffic intensity increases. With low traffic intensity, radio nodes are underutilized most of time. Therefore giving more paths (or serving cells) can provide only small gains. At high traffic intensity, some radio nodes are heavily loaded. Multipath transmission will improve load balancing and definitely bring larger gains over single path.

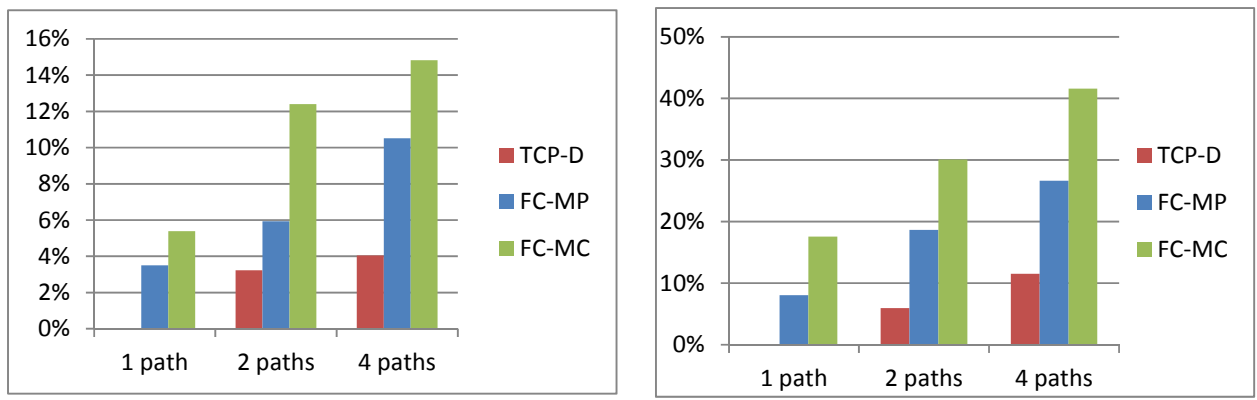

**Figure 4: Performance of protocols compared to the baseline 1-path TCP-D, low intensity traffic (left) and high intensity traffic (right)**

Our simulation results also verify that mobile users have better fairness [9] as they can experience good channel while travelling. Users have similar number of finished sessions in mobility scenario. This is opposite to the non-mobility case, where a significant disparity among users is observed.

## Impacts of Packet Forwarding Handover

In the previous section, we presented results with packet dropping handover. It is necessary to examine whether packet forwarding could improve performance of BE traffics. From results in Figure 5, the packet forwarding has negative impacts to TCP-D protocol, while gives similar performance as the case of packet dropping for FC-MP protocol. As analysed in Section 3, the packet forwarding handover and TCP retransmission may results in packet duplication and thus increases the file downloading time. On the other hand, fountain coded packets are independent. The forwarded packets may slightly help to improve the performance of FC-MP protocol as more FC packets are available when the users' channels have higher capacity. Anyway, the gain for FC-MP protocol is small, while packet forwarding requires additional complexity to the handover process.

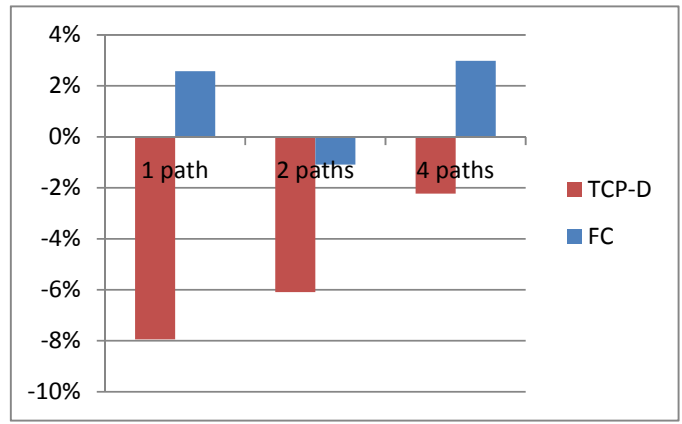

Figure 5: Relative performance change when packet forwarding is employed, high-intensity best-effort traffic.

## DN vs. Macro-cell Network

The DN has been shown to give a much higher capacity than the macro-cell network because of spatial reuse gain [15]. In this section, we examine the potential gain of DN in the mobility scenario, and with SDN and SDP implementation.

Simulation results are provided in Figure 6. The relative gain of DN (57 radio nodes in an area of 0.04 $km^2$) over macro-cell network (57 radio nodes in an area of 1 $km^2$) is significant for any mobile speed. When the user density increases, the gain also increases. Another observation is that the gain slightly reduces when users move faster. The performance of DN marginal changes as speed increases. On the other hand, high mobility can still improve performance of macro-cell UEs since UEs may quickly move out of poor signal regions within large coverage areas of macro-cells.

So with the proposed SDN-based and SD solutions, the network densification not only helps fixed users but also mobile users. Mobile users should be connected to small cells to enjoy higher data rate and better fairness.

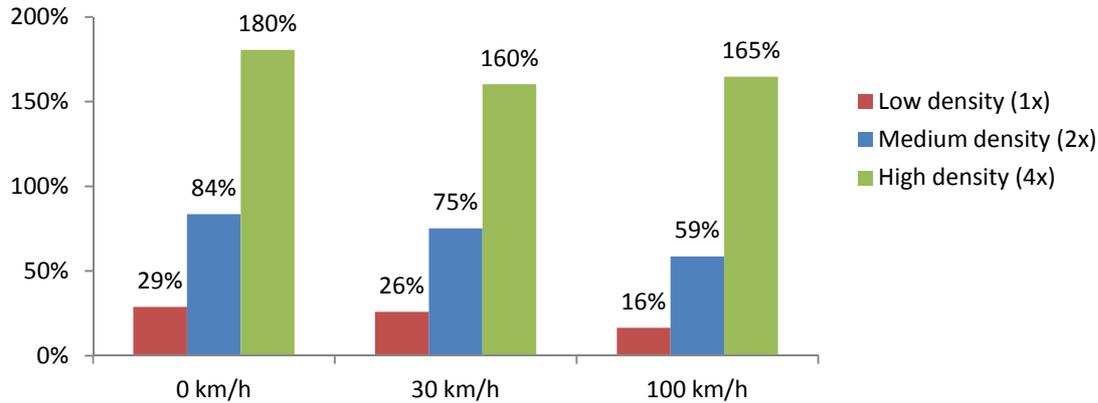

**Figure 6: Relative gain of dense small networks compared to macro-cell network.**

### 4.3   Performance of Transport Protocols for Real-Time Video

Simulation results are shown in Table 1. We compare the gain by using FC-MP protocol over a baseline scheme using single-path transmission and without fountain coding. Note that for the conventional UDP real-time video, using multipath TE could be inefficient because the packet delay in low capacity paths could be significant to degrade users' QoE of real-time video flows. From simulation results, the FC-MP protocol can provide remarkable gains at any mobile speeds.

The effectiveness of fixed rate fountain coding depends on the number of independently coded packets, which is dependent on the number of information bits. Since the size of I-frame can be large for high bit rates, it is possible to generate many coded packets for I-frames and distribute them to different radio nodes. In this way the transmission of I-frame can take advantages of macro diversity of multiple serving radio nodes, as well as time/frequency diversities when packets are scheduled in each radio nodes. However, the size of P-frames could be much smaller than that of I-frames. The number of coded packets for P-frame could be very small for fixed rate fountain codes, which may not help to exploit the macro diversity and time/frequency diversities. If FC-MP is not used to protect P-frames, the packet forwarding handover needs to be applied to P-frames to further improve the video decoding quality.

**Table 1: Performance gain of UDP and FC video transport protocols.**

| Speed (km/h) | 1-path UDP (kbit/s) | 4-path FC-MP (kbit/s) | Gain FC-MP vs. UDP |
|---|---|---|---|
| 0 | 700 | 1,100 | **57%** |
| 30 | 450 | 950 | **111%** |
| 100 | 300 | 500 | **67%** |

## 5   Improved Efficiency of Fountain Coding

In the previous section, we have shown how FC-MP protocol can significantly boost performance of SDN-based DN networks. This section presents techniques to improve the efficiency of FC-MP protocol in terms of redundancy and complexity.

**Packet Redundancy Reduction**

The FC-MP protocol performs significantly better than the TCP-D protocol because it creates redundant packets and the TE Optimizer properly distributes these packets to serving radio nodes. However, the TE decision may be suboptimal as channel state information may be inaccurate. Due to the imprecise rate allocation, the incoming rate of traffic flows to radio nodes could be higher than that can be handled by radio nodes. This leads to a waste of backhaul capacity and reduces the overall network performance, especially when the backhaul capacity is limited.

To mitigate this issue, we introduce a feedback mechanism so that radio nodes send back the buffer status to the v-u-SGW, where FC packets are generated (see Figure 1). Buffer status reports could be sent at higher frequency, e.g. every 0.1 second, while the TE is run at longer timescales, e.g. 0.5 second. The data rate of individual flows can be adjusted in between two TE runs according to its buffer size at radio nodes. For instance, if the buffer size is greater than a threshold, the incoming rate should be lessened. In our study, the number of redundant repair packets could be as high as 30% or more for a third of BE sessions in a scenario with two serving cells, 30 km/h, low traffic intensity. Using our technique, the redundancy can be cut down remarkably by 3.5 times.

**Complexity Reduction**

The encoding and decoding computational complexity of fountain code is only linear in the number of data symbols. Nevertheless, additional complexity such as buffer size at the source and sink should be taken into account. The NORM [7] and MBMS [8] file delivery protocols provides a measure to reduce the complexity by splitting large files into smaller segments. Fountain coding is applied to data segments with manageable encoding/decoding complexity and memory size of the user. This procedure is efficient in large-scaled multicast file distributions, where many users may loss different packets of the same data segments. In the unicast scenario, a user may lose a few packets in different segments. Then the error correction process needs to be applied for each segment, which may take a longer time to correct errors in all segments.

A more efficient error correction procedure, namely on-demand fountain coding (OD-FC), is to just apply fountain coding for missing data packets only. The traffic sink reports missing packets upon received a certain number of packets. The source and sink then may remove received packets from transmit and receive buffers, respectively, to save memory. If the number of missing packets is small, it may be impossible to generate many coded packets. One solution is to encode the missing packets and some lastly received packets so that the total number of packets entering the FC encoder will be enough to generate a large amount of coded packets. The performance of OD-FC protocol is verified by extensive simulations. The OD-FC protocol performs as well as FC-MP protocol, while the encoding/decoding complexity can be reduced extensively.

# 6 Conclusion

We presented a system architect and software-defined protocol concept to support mobile users in DN networks. Virtual user-specific serving gateway is dynamically selected to provide traffic engineering and SDP functions for mobile users. In our SDP solutions, fountain coding is an enabler to realize the promising gain of multipath traffic engineering. The SDP will select functional entities to provide reliable transmission, depending on traffic types and capability of network nodes. For example, when serving BE flows, SDP will select FC-MP protocol in v-u-SGW and turn off layer-2 ARQ, layer-1 HARQ (Hybrid ARQ), and packet forwarding handover. For delay-sensitive packets like I-frame, SDP can select fixed-rate FC-MP protocol; and because of fixed-rate coding, packet forwarding handover is needed for fountain coded video frames. If FC is not selected for some unimportant video frame, HARQ and packet forwarding are needed to improve the video decoding quality. Necessary features to support mobility are summarized in Table 2.

With our proposed solution package, the mobile users can enjoy high throughput and better fairness in dense networks. The association with macro cells is not an optimal solution for mobile users in terms of throughput. We also show how efficient the multipath traffic engineering combined with fountain coding could outperform convention multicast solution, which suffers from overused backhaul and complicated multi-cell scheduling coordination.

Our work could be extended in different directions. For example, user location tracking in mobility could be improved by prediction. End-to-end file delivery delay could be shortened by exploiting knowledge of channel capacity in the next user locations.

Table 2: Useful functions to support mobility in DN

| Function | Best Effort | Delay-Sensitive | |
| --- | --- | --- | --- |
| | | I-frame | P-frame |
| Traffic Engineering | | ✓ | ✓ |
| Radio Coordination | if high load | ✓ | ✓ |
| Multipath Transmission | ✓ | ✓ | ✓ |
| HARQ | | | ✓ |
| FC | ✓ | ✓ | Not efficient for small data rates |
| Handover Packet forwarding | | ✓ | if FC not used and if better video decoding quality required |
| Radio scheduler | Max-rate scheduling for mobile users | | |